\documentclass[12pt]{article}
\usepackage{a4wide}
\usepackage{graphicx}
\newcommand{\be}{\begin{equation}}
\newcommand{\ee}{\end{equation}}
\newcommand{\bea}{\begin{eqnarray}}
\newcommand{\eea}{\end{eqnarray}}
\newcounter{saveeqn}
\newcounter{App} 
\newcommand{\app}{%
\stepcounter{App}%
\setcounter{saveeqn}{\value{equation}}%
\setcounter{equation}{0}%
\renewcommand{\theequation}{\Alph{App}\arabic{equation}} }
\newcommand{\appende}{%
\setcounter{equation}{\value{saveeqn}}%
\renewcommand{\theequation}{\arabic{equation}}  }     
\begin{document}
\begin{flushright}
TTP02-16
\end{flushright}
\vspace{0.5cm}

\thispagestyle{empty}
\begin{center}
{\Large\bf Gluonic Penguins in  $B \to \pi \pi$ \\
from QCD Light-Cone Sum Rules}
\vskip 1.5true cm

{\large\bf
Alexander Khodjamirian\,$^{a,*)}$, 
Thomas Mannel\,$^{a}$, Piotr Urban\,$^{a,b}$} 
\\[1cm]~
\vskip 0.5true cm

{\it $^a$ Institut f\"ur Theoretische Teilchenphysik, Universit\"at
Karlsruhe, \\ D-76128 Karlsruhe, Germany\\
$^b$ Henryk Niewodnicza\'nski Institute of Nuclear Physics\\
Kawiory 26a, 30-055 Krak\'ow, Poland } 

\end{center}

\vskip 1.0true cm

\begin{abstract}
\noindent
The $B\to \pi\pi$ hadronic matrix element 
of the chromomagnetic dipole operator $O_{8g}$
(gluonic penguin) is calculated using the QCD 
light-cone sum rule approach. The resulting sum rule for 
$\langle \pi\pi |O_{8g}|B\rangle $ 
contains, in addition to the $O(\alpha_s)$ part induced by 
hard gluon exchanges, a contribution due to  
soft gluons.  
We find that in the limit $m_b\to \infty$ the soft-gluon contribution
is suppressed as a second power of $1/m_b$
with respect to the leading-order factorizable
 $B\to \pi\pi$ amplitude, whereas the hard-gluon contribution has 
only an $\alpha_s$ suppression. Nevertheless, 
at  finite $m_b$, soft 
and hard effects of the gluonic penguin 
in $B\to \pi\pi$ are of the same order.
Our result indicates that soft contributions 
are indispensable for an accurate counting 
of nonfactorizable effects in charmless $B$ decays.
On the phenomenological side we predict that the 
impact of gluonic penguins on $\bar{B}^0_d\to \pi^+\pi^-$  
is very small, 
but is noticeable for $\bar{B}^0_d\to \pi^0\pi^0$. 
\end{abstract}

\vspace*{\fill}

\noindent $^{*)}${\small \it On leave from 
Yerevan Physics Institute, 375036 Yerevan, Armenia } \\
\newpage
\section{Introduction}

Intensive experimental studies of 
$B\to \pi\pi$  and other charmless two-body hadronic 
$B$ decays are currently under way at $B$ factories ~\cite{Bpipiexp}.
The major goal is to complement the already observed 
CP-violation in the $B\to J/\psi K$ channel
by measuring other CP-violating effects 
and quantifying them 
in terms of the angles $\alpha$ and $\gamma$ 
of the CKM unitarity triangle (for a recent comprehensive review,
see, e.g. Ref.~\cite{Fleischer}). The  extraction of these fundamental 
parameters  from experimental data on $B\to \pi\pi, K\pi, K\bar{K}$ 
demands  reliable theoretical estimates of 
hadronic matrix elements of the  
effective weak Hamiltonian between the initial $B$ and the final 
two-light-meson states. These matrix elements are not yet 
accessible in lattice QCD, and thus other QCD methods of their calculation 
are actively being developed.

The factorization ansatz in hadronic two-body 
$B$ decays has recently been put on a more solid ground.
It has been shown \cite{BBNS}, to $O(\alpha_s)$, 
that in the limit $m_b\to \infty$ 
the amplitude for $B\to \pi\pi$ factorizes
into a product of the $B\to \pi$ form factor 
and the pion decay constant.
The nonfactorizable effects due to hard gluon exchanges 
are directly calculated combining perturbative QCD 
with certain nonperturbative inputs. The latter include the 
$B\to \pi$ form factor and  the light-cone distribution
amplitudes (DA) of pions and $B$ meson. 
According to QCD factorization, 
various soft nonfactorizable contributions
to $B\to \pi\pi$, such as exchanges of 
low virtuality gluons between the ``emission'' pion
and the remaining $B\to \pi$ part of the process, 
vanish in the $m_b \to \infty$ limit. 
An open phenomenological 
problem for QCD factorization is to obtain quantitative estimates of 
soft nonfactorizable effects at a finite, physical $b$ quark mass. 
Among these effects,  
the contributions of current-current operators  
in the penguin topology (e.g., the ``charming penguins'') could be 
important, as was argued in Ref.~\cite{charmpeng}.

Recently, a new method to calculate  the $B\to \pi\pi$ hadronic 
matrix elements from QCD 
light-cone sum rules (LCSR) \cite{lcsr,BF1,CZ} has been suggested 
by one of us \cite{AK}.
The main advantage of  LCSR is that 
the hadronic matrix elements of heavy-to-light 
transitions can be calculated 
including simultaneously soft and hard effects.
Thus, the $B\to \pi$ form factor 
including both soft (end-point) contributions and hard-gluon exchanges 
was obtained from LCSR \cite{Bpi,KRWY,BBB,BallZ}
providing the main input for the factorizable $B\to \pi\pi$ amplitude. 
Furthermore, in Ref.~\cite{AK} it was shown that with LCSR 
one achieves a quantitative control over $1/m_b$ 
effects of soft nonfactorizable gluons in $B\to \pi\pi$.    
The soft-gluon effect in the matrix element of the current-current 
operator  in the emission topology 
turns out to be of the same  size as the 
hard-gluon nonfactorizable contributions obtained from
QCD factorization. 
It is important therefore to investigate other 
nonfactorizable effects, related to the penguin and dipole  
operators and/or to non-emission topologies for the current-current operators.
Especially interesting is to compare the size of soft- and hard-gluon  
contributions.

To continue a systematic study in this direction 
we apply in this paper the LCSR method 
to the $B\to \pi\pi$ hadronic matrix element of 
the chromomagnetic dipole operator 
$O_{8g}$ (gluonic penguin).  
Apart from being phenomenologically 
interesting by itself, the $\langle \pi\pi| O_{8g}| B\rangle$ 
matrix element provides a very useful 
study case for the LCSR approach. Indeed, as we shall see, the 
hard-gluon contribution is given by one-loop diagrams
and is therefore relatively simple. 
The soft-gluon effect corresponding to a tree-level diagram 
is also easily calculable.
One is therefore able to estimate both hard and soft 
gluonic-penguin contributions using one and the same method and input.

The paper is organized as follows. In section 2
we introduce the relevant three-point correlation function 
and calculate it using operator-product expansion
(OPE) near the light-cone and taking into 
account both hard- ($O(\alpha_s)$) and 
soft-gluon contributions. 
In section 3, following the procedure described in Ref.~\cite{AK}
we match the result of this calculation to 
the dispersion relations in $\pi$ and $B$ channels
combined with  quark-hadron duality, and obtain the 
sum rule for the $\langle \pi\pi|O_{8g}|B\rangle$  
matrix element. In section 4 we investigate the $m_b\to \infty$ limit
of the sum rule. Section 5 contains the numerical analysis. 
Furthermore, in section 6 our prediction  is discussed
from the phenomenological point of view and compared, in section 7,
with the corresponding result of QCD factorization.
We conclude in section 8. The appendix contains 
some useful formulae: the  decompositions of
the vacuum-to-pion  matrix elements
in terms of the light-cone DA of various twist. 

\section{Correlation function}

Following Ref.~\cite{AK} we start with defining a 
vacuum-pion correlation function:
\be
F_\alpha (p,q,k)=
-\int d^4x\, e^{-i(p-q)x} \int d^4y\, e^{i(p-k)y}
\langle\, 0 \mid T\{j^{(\pi)}_{\alpha 5}(y)O_{8g}(0)j^{(B)}_5(x)\}\mid \pi^-(q)
\rangle\,,
\label{corr}
\ee
where the quark currents $j^{(\pi)}_{\alpha5}=\bar{u}\gamma_\alpha \gamma_5 d $ 
and $j^{(B)}_5 = m_b \bar{b}i \gamma_5 d$ interpolate 
$\pi$ and $B$ mesons, respectively, and 
the quark-gluon chromomagnetic dipole operator 
$O_{8g}$ relevant for $B\to \pi\pi $ has the following 
standard expression:
\be
O_{8g}=\frac{m_b}{8\pi^2}\bar{d}\sigma^{\mu\nu}(1+\gamma_5)
\frac{\lambda^a}2 g_sG^a_{\mu\nu} b\,.
\label{peng}
\ee
This operator enters the $\Delta B =1 $ effective weak Hamiltonian
\be
H_W= \frac{G_F}{\sqrt{2}}
\left\{ \lambda_u\left[\left( c_1(\mu)+ \frac{c_2(\mu)}{3}\right)O_1(\mu) +
  2c_2(\mu)\widetilde{O}_1(\mu)\right]
+ ... +\lambda_t c_{8g}(\mu)O_{8g}(\mu)\right\}\,, 
\label{H}
\ee
where $\lambda_q=V_{qb}V^*_{qd}$, $q=u,t$ are the combinations of CKM parameters, 
and we have singled out the relevant current-current operators:    
\be
O_1=(\bar{d}\Gamma_\mu u)(\bar{u}\Gamma^\mu b)\,,~~~
\widetilde{O}_1=(\bar{d}\Gamma_\mu \frac{\lambda^a}2u)(\bar{u}\Gamma^\mu
\frac{\lambda^a}2 b)\,,
\label{o1}
\ee 
denoting by ellipses all remaining operators which
we do not consider in this paper. 
In Eqs.~(\ref{peng})-(\ref{o1}),
$\mbox{Tr}(\lambda^a\lambda^b) = 2\delta^{ab}$, $c_{1,2,8g}$ are the Wilson
coefficients, $\mu\sim m_b$ is the normalization scale and $\Gamma_\mu=\gamma_\mu(1-\gamma_5)$. 
Furthermore, the operator $O_2=(\bar{u}\Gamma_\mu u)(\bar{d}\Gamma^\mu b)\,$ 
has been Fierz transformed: $O_2= O_1/3+ 2\widetilde{O}_1$.

The correlator (\ref{corr}) is a function of 
three independent momenta, which are chosen to be $q$, $p-k$, $k$,
and, for simplicity, $p^2=k^2=0$. 
Similar to Ref.~\cite{AK}, we introduce the unphysical external momentum 
$k$ flowing into the $O_{8g}$ vertex. At $k\neq 0$, the momentum $p-q$ 
in the $B$ channel is independent of the total momentum $P=p-q-k$ of the 
light-quark state formed after the $b$ quark decay. That 
will allow us later to avoid contributions of the ``parasitic'' light-quark 
states in the dispersion relation in the variable $(p-q)^2$.
In the final sum rule $k$ vanishes in the ground-state $B\to \pi\pi$ 
contribution. 
In Eq.(\ref{corr}) the pion is on shell, $q^2=m_\pi^2$. 
We will work in the chiral limit
and set $m_\pi=0$ everywhere, except in the enhanced 
combination $\mu_\pi=m_\pi^2/(m_u+m_d)$.
The Lorentz-decomposition of the correlation function (\ref{corr})
contains four invariant amplitudes:
\be
F_\alpha= 
(p-k)_\alpha F 
+ q_\alpha \widetilde{F}_1  
+ k_\alpha\widetilde{F}_2 
+ \epsilon_{\alpha\beta\lambda\rho}q^\beta p^\lambda k^\rho
\widetilde{F}_3\,,
\label{decompos}
\ee
depending
on three kinematical invariants $(p-k)^2$, $(p-q)^2$ and 
$P^2=(p-q-k)^2$. In what follows only the amplitude $F$ is needed.    

In the region of large virtualities: $(p-k)^2,(p-q)^2,P^2<0$,
$|(p-k)^2|,|(p-q)^2|,|P^2|\gg \Lambda_{QCD}^2$,
the correlation function (\ref{corr}) is calculated  
using light-cone OPE, in a form of the sum of short-distance coefficient functions (hard amplitudes) convoluted with the pion DA of 
growing twist and multiplicity. 
Only a few first terms of this expansion are
usually retained, the components with higher twist/multiplicity are
neglected being suppressed by  inverse powers of large virtualities.

The gluon emitted from the vertex $O_{8g}$ in the correlation function 
contributes in two different ways. First, a highly virtual (hard) gluon
is absorbed by one of the quark lines at a short light-cone separation
from the emission point. The corresponding one-loop diagrams 
(see Fig.~1) are then calculated perturbatively, as 
an $O(\alpha_s)$ part of the 
hard amplitude. The second possibility is that the emitted gluon  
together with the quark-antiquark pair forms
the quark-antiquark-gluon DA of the pion (see Fig.~2a). We will call 
these low-virtuality gluons {\em soft}. 
Thus, in the correlation function the hard- and soft-gluon 
contributions belong to different terms of the light-cone OPE 
and  can be calculated separately.
In what follows, we take into account 
all $O(\alpha_s)$ hard-gluon effects of twist 2 and 3, and, in addition 
the effects of  quark-antiquark-gluon DA of twist 3 and 4 in zeroth order 
in $\alpha_s$. 
This approximation corresponds to the accuracy of the 
light-cone expansion adopted for the 
LCSR calculation of the $B\to \pi$ form factor.

The higher twist terms 
corresponding to the soft-gluon contributions are 
subleading in the light-cone OPE, being suppressed
by inverse powers of large virtualities.
However, in the correlation function (\ref{corr}) the lowest twist 
contribution is of $O(\alpha_s)$, therefore both hard and soft contributions 
can be equally important. We may again refer to the LCSR for 
the $B\to \pi$ form factor where the $O(\alpha_s)$ twist 2 and 
the zeroth-order in $\alpha_s$ twist 4  contributions are 
at the same level numerically.

In the following three subsections the  
calculation of the various contributions to the correlation function
(\ref{corr}) is presented. 

\subsection{Hard gluon contribution}

We begin with calculating the leading $O(\alpha_s)$ contribution 
to the correlator (\ref{corr}) generated by hard gluons.
To obtain the relevant diagrams 
one has to contract the $b$-quark fields 
and, in addition, the $d$- and $\bar{d}-$quark  
fields from the operators $j^{(\pi)}_{\alpha5}$
and $O_{8g}$, respectively, inserting
the free-quark propagators for both contractions.
\begin{figure}[tb]
\begin{center}
\includegraphics[width=0.9\textwidth]{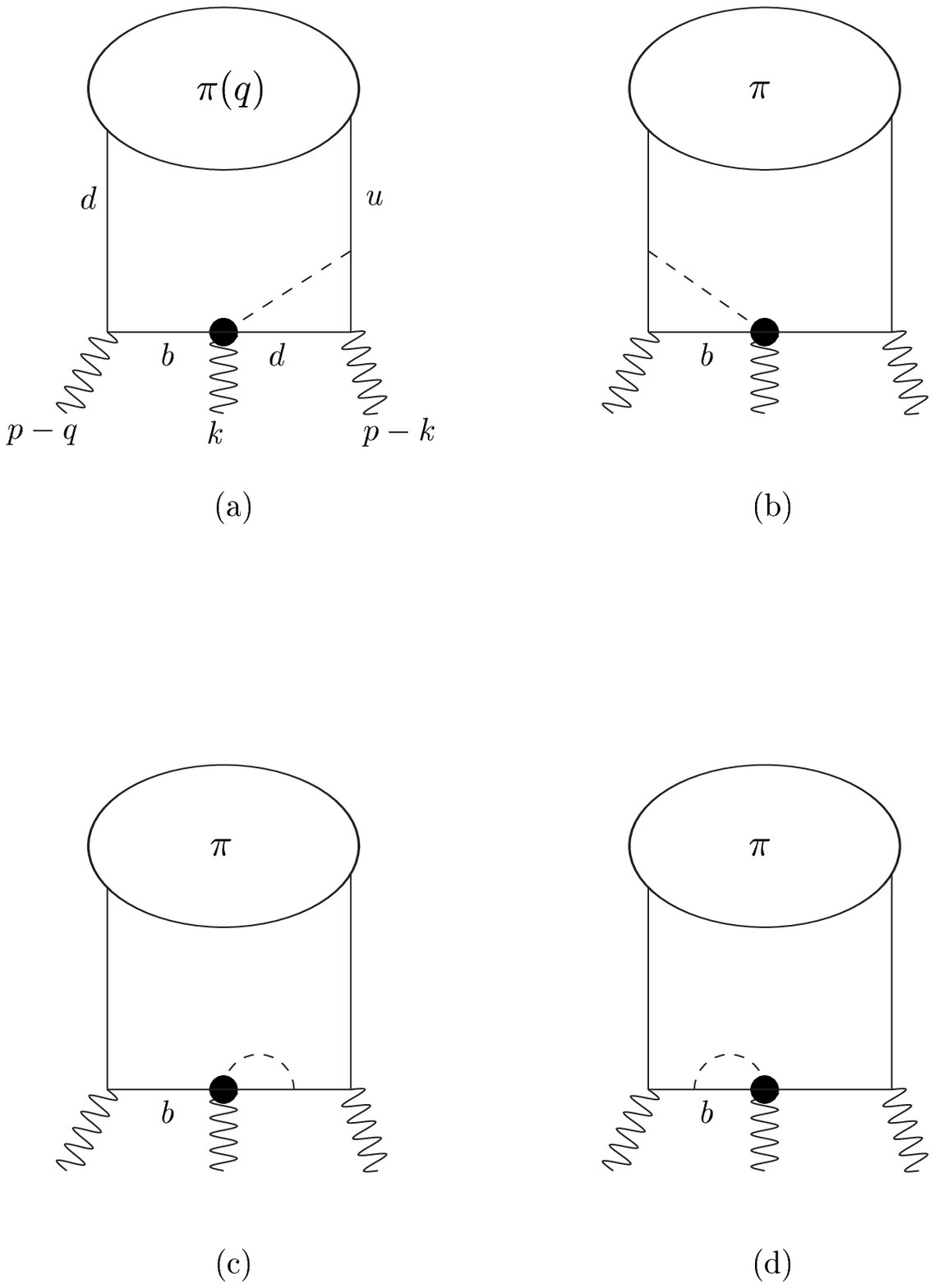}
\end{center}
\caption{
{\it Diagrams corresponding to the perturbative $O(\alpha_s)$ contributions
to the correlation function (\ref{corr}). 
Solid, dashed, wavy lines and ovals represent quarks, gluons,  
external currents and pseudoscalar meson DA, respectively.
Thick points indicate the gluonic penguin vertex. }
\label{fig1}}
\end{figure}
The remaining on-shell $\bar{u} $ and $d$ 
fields form the quark-antiquark DA of the pion.
The decomposition of the relevant vacuum-pion 
matrix elements in terms of DA of the lowest twists 
2 and 3 is presented in the appendix. Twist 4 components
are neglected, because the $O(\alpha_s)$  twist-4 effects 
are beyond the accuracy of our calculation. 
The gluon emitted by $O_{8g}$ is absorbed by one of the four quark lines
leading to the four diagrams shown in Fig.~1. 
Note that we disregard the possibility to 
contract the $\bar{d}$ quark from $O_{8g}$ 
with the $d$ quark  from $j_5^{(B)}$. That type of contraction 
leads to ``penguin-annihilation'' diagrams where 
the heavy-light loop is connected by a single gluon with the 
pion part of the correlation function. An additional gluon 
has to be added in this case to obey color conservation, making 
this contribution either $O(\alpha_s^2)$,
with both gluons hard, or  $O(\alpha_s)$, with one hard gluon and one 
soft gluon, the latter entering the pion DA. Both effects are 
beyond the adopted approximation for the correlation function
and presumably very small. 

The one-loop diagrams are calculated employing dimensional
regularization. 
Among other tools we used the FORM program \cite{FORM}. 
The invariant amplitude $F$ is obtained by 
taking the coefficient at
$(p-k)_\alpha$. 
The result for $F$  is ultraviolet divergent; however, 
this divergence does not play any role in the resulting sum rule 
because the $1/\epsilon$  terms  vanish after Borel transformations. 
On the other hand, we find that 
the amplitude $F$  does not contain infrared-collinear 
divergences. This is in accordance with the fact that the perturbative 
expansion for the correlation function starts, in
twists 2 and 3, with $O(\alpha_s)$.
If one takes into account an additional perturbative gluon correction  
in the correlation function, the resulting two-loop diagrams 
will yield infrared-collinear divergences which 
should be absorbed by the evolution of the pion DA, as in 
the case of the $O(\alpha_s)$ correction to the 
LCSR for the $B\to \pi$ form factor \cite{KRWY,BBB}.  

According to the procedure of  
the sum rule derivation explained in Ref.~\cite{AK} 
we have to retain in the final answer for the amplitude $F$ 
only those terms which, in the limit of large $|P^2|\sim m_B^2$, 
have a nonvanishing double imaginary part 
in the variables $(p-k)^2$ and $ (p-q)^2 $, taken in the
duality region $\;0<(p-k)^2<s_0^\pi$,
$m_b^2<(p-q)^2<s_0^B$. We find that 
only two diagrams, in Fig.~1a,b, 
contain nonvanishing terms. Retaining  
only these terms we obtain the following  analytical expression 
for the hard-gluon contribution to the amplitude $F$: 
\be
F^{hard}(s_1,s_2,P^2)=F^{tw2}(s_1,s_2,P^2)+F^{tw3}(s_1,s_2,P^2)\,,
\ee
where the twist 2 and 3 parts are, respectively 
\bea
F^{tw2}(s_1,s_2,P^2)= \frac{\alpha_s C_F m_b^2 f_\pi}{16\pi^3}
\int\limits_0^1 du~\varphi_\pi(u)
\nonumber
\\
\times \left\{ 
\frac{s_2}{m_b^2-\bar{u}s_2}
\mbox{ln}\left(\frac{-s_1}{\mu^2} \right)+
\frac{(s_1+s_2-P^2)(m_b^2-s_2)^2}{u(-\bar{u}P^2-us_1)s_2^2}
\mbox{ln}\left (\frac{m_b^2-s_2}{m_b^2}\right)
\right \}\,,
\eea
and 
\bea
F^{tw3}(s_1,s_2,P^2)= \frac{\alpha_s C_F m_b^3 f_\pi\mu_\pi}{32\pi^3}
\Bigg[\int\limits_0^1 du~\varphi_p(u)
\Bigg\{ \frac{3}{m_b^2-\bar{u}s_2}
\mbox{ln}\left(\frac{-s_1}{\mu^2} \right)
\nonumber
\\
+\frac{(s_1-3s_2-P^2)(m_b^2-s_2)}{(-\bar{u}P^2-us_1)s_2^2}
\mbox{ln}\left (\frac{m_b^2-s_2}{m_b^2}\right)
\Bigg \}
\nonumber
\\
+\frac{1}{6}
\int\limits_0^1 du~\varphi_{\sigma}(u)
\Bigg\{\left(\frac{2}{\bar{u}(m_b^2-\bar{u}s_2)}
+\frac{s_2}{(m_b^2-\bar{u}s_2)^2}\right)
\mbox{ln}\left(\frac{-s_1}{\mu^2}\right)
\nonumber
\\
-\frac{(m_b^2-s_2)}{s_2^2}\left(\frac{2(s_1+s_2-P^2)}{u(-\bar{u}P^2-us_1)}+
\frac{(s_1-P^2)(s_1+5s_2-P^2)}{(-\bar{u}P^2-us_1)^2}\right)
\mbox{ln}\left(\frac{m_b^2-s_2}{m_b^2}\right)
\Bigg\}\Bigg]\,.
\label{hard}
\eea
In the above, we denote $s_1=(p-k)^2$, $s_2=(p-q)^2$ and $\bar{u}=1-u$ for brevity. 
The terms proportional to $\mbox{ln}(-s_1/\mu^2)$ and 
$\mbox{ln}((m_b^2-s_2)/m_b^2)$ originate from the diagrams
in Fig.~1a and 1b, respectively.
Note that the logarithmic dependence on the 
dimensional regularization scale $\mu$    
is unimportant because it vanishes after Borel transformation
in the variable $s_{1}$.

\subsection{Soft gluon contribution}

In order to reach the same
accuracy in the light-cone expansion of the correlation function
(\ref{corr}) as in the calculation of the $B\to \pi$ form factor 
one has to take into account the contributions 
of  quark-antiquark-gluon three-particle DA. Being 
a regular part of the light-cone expansion, these contributions 
correspond to a nonperturbative effect  of zeroth order  in $\alpha_s$
which is qualitatively different from hard-gluon exchanges. 
The on-shell gluon field emitted by the penguin operator
is directly  absorbed in the three-particle pion DA
yielding a single tree-level diagram shown in Fig.~2a. 
The corresponding expression is obtained 
by contracting $b$ and $d$ quark fields and using 
the decomposition of the  
$\langle 0|\bar{u}(y)G_{\mu\nu}(0) d(x)|\pi\rangle $
matrix element in terms of twist 3 and 4 quark-antiquark-gluon DA given in the appendix.   
We find that the contribution of twist 3 DA vanishes
in the amplitude $F$ and the answer contains only 
twist 4 DA :
\bea
F^{soft}(s_1,s_2,P^2)=
\frac{m_b^2 f_\pi}{8\pi^2}\int \frac{{\cal D}\alpha_i}{
(m_b^2-(1-\alpha_1)s_2)(-\alpha_2P^2-(1-\alpha_2)s_1)}
\nonumber
\\
\times
\{2(s_1-s_2-P^2)[\varphi_\perp(\alpha_i) +\widetilde{\varphi}_\perp
(\alpha_i))]+(s_1+s_2-P^2)[\varphi_\parallel(\alpha_i) +
\widetilde{\varphi}_\parallel(\alpha_i)] \}\,,
\label{soft}
\eea
where $\alpha_i\equiv \{\alpha_1,\alpha_2,\alpha_3\}$ and 
${\cal D} \alpha_i = d\alpha_1d\alpha_2d\alpha_3
\delta \left(1\!-\! \alpha_1\! -\! \alpha_2 \!-\! \alpha_3 \right)$. 
\begin{figure}[htb]
\begin{center}
\includegraphics[width=1.0\textwidth]{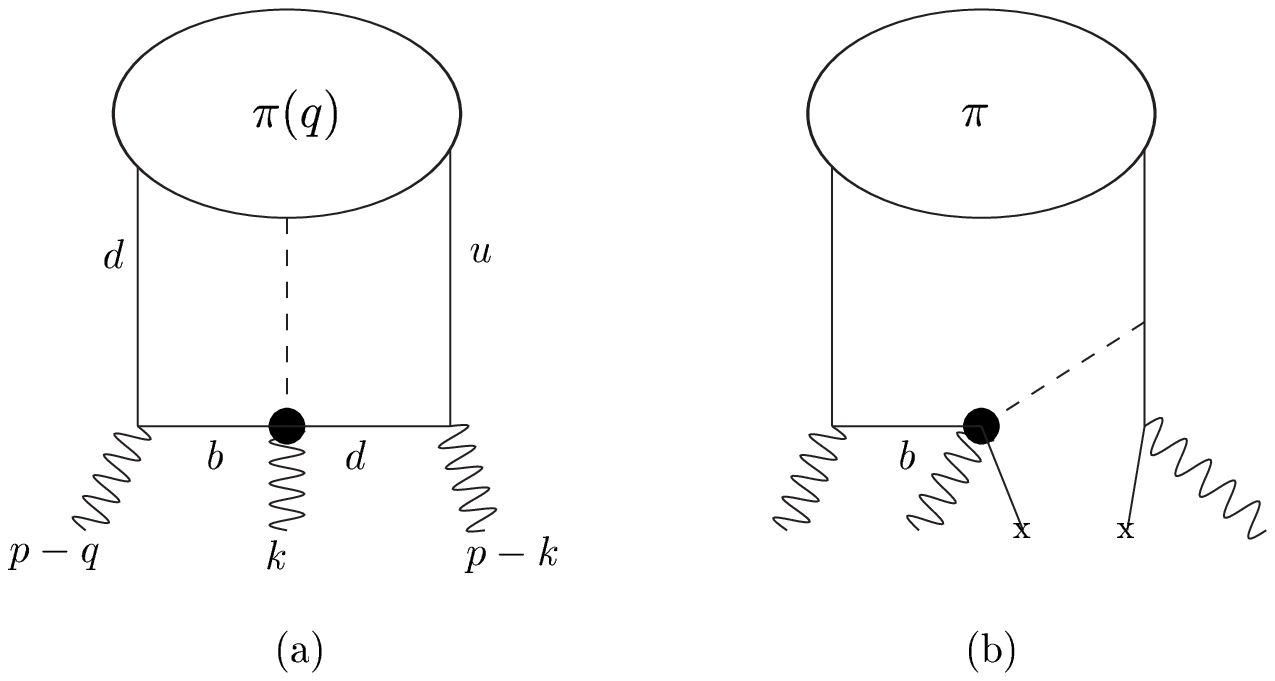}
\end{center}
\caption{
{\it Diagrams corresponding to the soft-gluon (a) and factorizable 4-quark
  (b) contributions
to the correlation function (\ref{corr})}. 
\label{fig:fig2}}
\end{figure}

\subsection{Factorizable 4-quark contribution }

Continuing the light-cone expansion of the correlation function 
(\ref{corr}) one encounters also four-particle Fock states. 
In particular, four-quark ($\bar{q}q\bar{q}q$) states  emerge when
$b$ quark and hard gluon propagate at short light-cone separations  and 
four light-quark fields are on-shell, yielding $O(\alpha_s)$ 
contribution. One of the corresponding diagrams is shown in Fig.~2b.
The Fock states with $\geq$ 3 multiplicity contribute in two different ways. 
First, various multiparticle DA of higher twist can be formed. 
Usually, in LCSR applications 
their contributions are expected to be strongly suppressed
(see, e.g. a more detailed discussion in Ref.~\cite{BKM}) and are therefore 
neglected, also in our analysis. Second, 
the configurations of multiple quark and gluon fields  
sandwiched between the vacuum and pion state, 
can in general be splitted into  a product of two operators
of lower multiplicity, one of them  being vacuum  averaged,
and the other one forming a vacuum-pion matrix element. In order
to avoid double counting, these factorizable contributions have to be
subtracted from genuine multiparticle DA.
In our case, the four-quark configurations emerging from the correlator (\ref{corr})
contain a factorizable piece, a product of vacuum-averaged quark-antiquark fields
and a low-twist (2,3) quark-antiquark DA. We will take the four-quark factorizable contributions into account approximating the
vacuum average of quark and antiquark fields with the standard quark
condensate of dimension 3. Combined with the twist 2 and 3 DA, these 
contributions are effectively of twist 5 and 6 respectively, and  
are analogous to the factorizable twist 6 terms taken into account in the
LCSR for  the pion electromagnetic form factor derived in
Ref.~\cite{BKM}. Other factorizable multiparticle 
contributions, such as $\bar{q}qGG$ or $\bar{q}q\bar{q}qG$ 
which can be parameterized in terms of dimension 4 (gluon)
and 5 (quark-gluon) condensates are neglected. 
The purpose of taking into account 
the quark-condensate contribution is twofold. First, it is enhanced 
by the numerically large parameter $\mu_\pi$,
therefore the size of the quark-condensate contribution to LCSR 
serves as an upper limit for all neglected multiparticle contributions. 
Second, the QCD factorization answer for the 
gluonic penguin amplitude contains a certain counterpart of the quark-antiquark
condensate term, with which our result will be compared.

Turning to the actual calculation, we find that 
only one diagram (Fig.~2b) provides nonvanishing contributions to 
the sum rule.
The short-distance part of this diagram consists 
of $b$- and $u$- quark propagators. 
The momenta of the vacuum averaged $\bar{d}d$-quarks are neglected 
in the adopted condensate approximation and the vacuum averaging 
is done in a usual way: $\langle 0|\bar{d^i_\xi}d^j_\omega|0\rangle=
\delta^{ij}\delta_{\omega\xi}\langle\bar{q}q \rangle/12 $,
where $\langle\bar{q}q \rangle$ is the quark condensate
density, $q=u,d$, and $i,j$, and $\xi,\omega$ are the colour and spinor
indices, respectively. The diagram in Fig.~2b yields:  
\bea
&&F^{\langle \bar q q \rangle}(s_1,s_2,P^2)= 
-\frac{\alpha_s C_F m_b^3 \langle \bar{q}q\rangle f_\pi}{12\pi}
\int\limits_0^1 \frac{du}{s_1(-\bar{u}P^2-us_1)(m_b^2-\bar{u}s_2)}
\nonumber
\\
&&
\times \Bigg\{2(s_1-P^2)\varphi_\pi(u)+
\frac{\mu_\pi}{2m_b}\left[3\bar{u}(s_1-s_2-P^2)-2s_1\right ]
\varphi_p(u) 
\nonumber
\\
&&
+\frac{\mu_\pi}{12m_b}\Bigg[ 15(s_1-P^2)+9s_2+
\frac{(s_1-P^2)(4s_1-3\bar{u}s_2)-5\bar{u}(s_1-P^2)^2+6s_1s_2}{-\bar{u}P^2-us_1}
 \nonumber
\\
&&
+\frac{5\bar{u}s_2(s_1-P^2)+3\bar{u}s_2^2+2s_1s_2}{m_b^2-\bar{u}s_2}
\Bigg]\varphi_\sigma(u)
\Bigg\}\,,
\label{cond}
\eea
where only the terms which contribute to the sum rule are retained. 
Adding together the contributions presented 
in Eqs.~(\ref{hard})-(\ref{cond}), we obtain our final
result for the amplitude $F$:
\be
F(s_1,s_2,P^2)=F^{hard}(s_1,s_2,P^2) + F^{soft}(s_1,s_2,P^2) + 
F^{\langle \bar{q}q \rangle}(s_1,s_2,P^2)
\label{sum}
\ee 
which will be used to derive LCSR.

\section{Dispersion relation and sum rule}

The procedure of deriving the sum rule for the 
$B\to \pi\pi$ matrix element from the correlation
function (\ref{corr}) is independent of the operator $O_{8g}$. Therefore
we refer to the paper \cite{AK} where this procedure was 
explained in detail for a generic case. 
For completeness, let us summarize the main steps of this derivation:
\begin{enumerate}
\item The calculated correlation function
(the amplitude $F$ in Eq.~(\ref{sum})) 
is matched to the dispersion relation in the pion channel, in the
variable $s_1=(p-k)^2$, at large $s_1<0$.

\item The Borel transformation, $s_1 \to M_1^2$,
is performed and quark-hadron duality is used to approximate the
contribution of excited pseudoscalar and axial states (with an effective
threshold $s_0^\pi$) yielding, at large spacelike $P^2\sim -m_b^2$,
an  expression for the matrix element 
of the operator product $O_{8g}j_B$ between pion states.

\item The matrix element $\langle \pi| O_{8g}j_B|\pi\rangle$ 
is expanded in powers of the small ratio $s_0^\pi/(-P^2)$
retaining  the leading term of this expansion. In the resulting
expression  the analytic continuation from large spacelike $P^2$ 
to large timelike $P^2=m_B^2$ is performed. 
Importantly, in the adopted approximation for the correlation function 
the matrix element $\langle \pi\pi | O_{8g}j_B|0\rangle|_{P^2=m_B^2}$
has no complex phase, as directly follows from the calculation of the imaginary part in
$s_1$  of the relevant diagrams in Figs.~1,2. At $0<s_1<s_0^\pi$ 
the resulting expression  remains real at $P^2 \to m_B^2$.  
An imaginary part will appear if one adds hard gluon exchanges to the diagrams
in Figs.~1,2, proceeding beyond the adopted order in $\alpha_s$. 
Physically, the imaginary part of $\langle \pi\pi |
O_{8g}j_B|0\rangle|_{P^2=m_B^2}$ should be identified, within the
accuracy of the quark-hadron duality approximation, with the phase 
of the final-state rescattering of two pions. Our result indicates
the smallness of this phase.  

\item The expression obtained for $\langle \pi\pi | O_{8g}j_B|0\rangle$
is then equated to the dispersion relation in the $B$-channel, in the  variable $s_2= (p-q)^2$.
The auxiliary momentum $k$ vanishes
in the ground  state $B$-meson contribution. The latter contains 
the hadronic matrix element of $O_{8g}$       
between  $B$ and $\pi\pi$ states multiplied by the $B$-meson decay 
constant $f_B$.

\item The second Borel transformation, $s_2 \to M_2^2$, is applied to 
the resulting relation and quark-hadron duality is employed 
again, with the corresponding threshold  $s_0^B$, to approximate the 
contribution of  the excited $B$ states. 
Finally, the desired sum rule for the $\langle \pi\pi | O_{8g}|B\rangle$
matrix element is obtained.
\end{enumerate}

For convenience, we present the resulting LCSR 
in a form of the sum of three separate contributions: 
\be
A^{(O_{8g})}( \bar{B}^0_d \to \pi^+ \pi^-) 
\equiv \langle \pi^-(p)\pi^+(-q) |O_{8g} |\bar{B}^0_d(p-q) \rangle
=A^{(O_{8g})}_{hard}+A^{(O_{8g})}_{soft}+A^{(O_{8g})}_{\langle\bar{q}q\rangle}\,,
\label{lcsr}
\ee
where
\bea
A^{(O_{8g})}_{hard}\!\!\!\!\!&=&\!\!\!\!\! i\frac{\alpha_s C_F}{2\pi} m_b^2
\Bigg(\frac{1}{4\pi^2f_\pi}\int\limits_0^{s_0^\pi}
ds~e^{-s/M_1^2}\Bigg)
\Bigg(\frac{m_b^2f_\pi}{2m_B^2f_B}\int\limits_{u_0^B}^1 \frac{du}{u}~
e^{\,m_B^2/M_2^2-m_b^2/uM_2^2 }
\nonumber 
\\
&\times&\!\!\!\!\!\Bigg [\frac{\varphi_\pi(u)}{u}+\frac{\mu_\pi}{2m_b}
\Bigg\{3\varphi_p(u)
+\frac{\varphi_\sigma(u)}{3u}-\frac{\varphi_\sigma'(u)}{6}
\nonumber
\\
&-&\varphi_p(1)\bar{u}\left(1+
\frac{3m_b^2}{um_B^2}\right)
+\frac{\varphi'_\sigma(1)\bar{u}}{6}\left(\frac{5m_b^2}{ u
    m_B^2}-1\right)\Bigg\}
\Bigg]\Bigg),
\label{Ahard}
\\
A^{(O_{8g})}_{soft}\!\!\!\!\!&=&\!\!\!\! -im_b^2
\Bigg(\frac{1}{4\pi^2f_\pi}\int\limits_0^{s_0^\pi}
ds~e^{-s/M_1^2}\Bigg)
\Bigg(\frac{f_\pi}{m_B^2f_B}\int\limits_{u_0^B}^1 \frac{du}{u}~
e^{\,m_B^2/M_2^2-m_b^2/uM_2^2 }
\nonumber
\\
&\times&\left(1+\frac{m_b^2}{um_B^2}\right)\Big[\varphi_\perp(1-u,0,u)
+\widetilde{\varphi}_\perp(1-u,0,u)\Big] 
\Bigg ),
\label{Asoft}
\\
A^{(O_{8g})}_{\langle\bar{q}q\rangle}\!\!\!\!\!&=&\!\!\!\!i\frac{\alpha_s C_F}{3\pi}m_b^2
\Bigg(\frac{-\langle \bar{q}q\rangle}{f_\pi m_b}\Bigg)
\Bigg(\frac{m_b^2f_\pi}{2m_B^2f_B}\int\limits_{u_0^B}^1\frac{du}u
~e^{\,m_B^2/M_2^2-m_b^2/uM_2^2}
\nonumber
\\
&\times&\!\!\!\!\!
\Bigg[ \frac{\varphi_\pi(u)}{u}+\frac{\mu_\pi}{4m_b}\Bigg\{3\left(1+\frac{m_b^2}{u m_B^2}\right)\varphi_p(u)
+\left(5-\frac{3m_b^2}{um_B^2}\right)\!\!
\left(\frac{\varphi_\sigma (u)}{3u}
-\frac{\varphi'_{\sigma}(u)}{6}\right)\!\Bigg\}\!\Bigg]\!\Bigg)\,.
\label{Aqq}
\eea
In the above, $\varphi'_\sigma(u)= \partial\varphi_\sigma(u)/\partial u$, 
$u_0^B=m_b^2/s_0^B$ and $f_B$ is the $B$-meson 
decay constant defined as
$m_b\langle 0|\bar{q}i\gamma_5b|B\rangle =m_B^2f_B$.
Note that the first bracket in Eqs. (\ref{Ahard}) and (\ref{Asoft})
is approximately equal to $f_\pi$ if one uses 
the SVZ sum rule \cite{SVZ}:
\be
f_\pi^2=\frac{1}{4\pi^2}\int\limits_0^{s_0^\pi}ds~e^{-s/M_1^2}\left (1+\frac{\alpha_s}{\pi}\right)+
\frac{\langle 0|\alpha_s/\pi G_{\mu\nu}^aG^{a \mu\nu} |0 \rangle}{12M^2}
+\frac{176}{81M^4}\pi\alpha_s\langle \bar{q}q\rangle ^2\,,
\label{bracket}
\ee 
retaining only the leading-order quark-loop term.
The quark-condensate contribution to  Eq.~(\ref{bracket}) of the
form $(m_u+m_d)\langle \bar{q}q\rangle/M^2$ is absent in the chiral limit,
whereas the perturbative correction and the gluon and 4-quark condensate
contributions are beyond our approximation. 

The second and third lines in Eq.~(\ref{Ahard}) contain the contributions
of the diagrams in Fig.~1a and 1b, respectively. In the latter contribution only twist 3
DA remain, the twist 2 term proportional to the small ratio 
$s_0^\pi/m_B^2$ is neglected. For the same reason,  
in the soft-gluon contribution (\ref{Asoft}) 
only two DA, $\varphi_\perp$ and $\widetilde{\varphi}_\perp$ contribute.

To minimize theoretical uncertainties 
we will calculate the ratio 
of the gluonic penguin matrix element (\ref{lcsr})
to the factorizable $B\to \pi\pi$ hadronic amplitude. The latter, as shown 
in Ref.~\cite{AK}, simply coincides with the matrix element 
of the operator $O_1$ in the emission topology: 
\begin{equation}
A^{(O_1)}_E( \bar{B}^0_d \to \pi^+ \pi^-) 
\equiv \langle \pi^-(p)\pi^+(-q) |O_1 |\bar{B}^0_d(p-q) \rangle_E=
im_B^2f_\pi f^+_{B\pi}(0)\,,
\label{fact}
\end{equation}
if one neglects the $O(s_0^\pi/m_B^2)$ terms
and the two-gluon nonfactorizable exchanges
which are beyond the adopted approximation.
In the above, $f^+_{B\pi}(0)$ is the LCSR result for the $B\to \pi$ form factor
at zero momentum transfer. For completeness
we write down  this sum rule  \cite{Bpi,KRWY,BBB}:
\bea 
f^+_{B\pi}(0)&=&\frac{f_{\pi}m_b^2}{2m_B^2f_B}
\int\limits_{u_0^B}^{1}\frac{du}{u}
e^{m_B^2/M_2^2-m_b^2/uM_2^2}
\Bigg( \varphi_\pi(u)
+ \frac{\mu_\pi}{m_b}
\Bigg[u\varphi_{p}(u) + \frac{
\varphi_{\sigma }(u)}{3}-\frac{u\varphi_\sigma'(u)}{6}\Bigg]
\Bigg)
\nonumber
\\
&+&O(q\bar{q}\mbox{tw4})+O(q\bar{q}G ) + O(\alpha_s)\,,
\label{Bpilcsr}
\eea 
where, for brevity, the subleading contributions 
of the twist-4 quark-antiquark DA, twist-3,4 quark-antiquark-gluon DA and the
$O(\alpha_s)$ correction to the twist 2 term  
are not shown explicitly but taken into account in the numerical 
analysis. The complete expression for LCSR (\ref{Bpilcsr})
can be found, e.g. in the review \cite{KR}. The recently calculated \cite{BallZ}
$O(\alpha_s)$ correction to the twist-3 part of the sum rule 
(\ref{Bpilcsr}) is very small and we neglect it here.

\section{Heavy quark mass limit}
The sum rule  (\ref{lcsr}) is derived in full QCD, with the finite $b$-quark mass.  
To obtain the heavy-quark mass limit of the matrix element 
$A^{(O_{8g})}( \bar{B}^0_d \to \pi^+ \pi^-)$, 
we  substitute  in LCSR the expansions of all  
$m_b$-dependent quantities:
\be
m_B = m_b+\bar{\Lambda}~,~~~ s_0^B = m_b^2 + 2m_b\omega_0 ~,
~~~M_2^2= 2m_b\tau~,~~~f_B = \hat{f}_B/\sqrt{m_b}, 
\label{hql1}
\ee
where 
$\bar{\Lambda}$, $\omega_0$, $\tau$, and $\hat{f}_B$ are
the parameters independent of the heavy-mass scale. One also has  to take into account 
that, at $m_b\to \infty$, $u_B^0\to 1-\omega_0/m_b$ and 
only the end-point regions of the momentum fraction $u$ contribute in
the integrals in  Eqs.~(\ref{Ahard})-(\ref{Aqq}). Using a convenient 
integration variable $\rho=\bar{u}m_b$, and taking into account
the end-point behavior of DA, the following substitutions have to be done
in these integrals, to the leading order in $O(1/m_b)$ :
\bea
\int\limits _{u_B^0}^1 du =(1/m_b)\int\limits_0^{2\omega_0}d\rho\,,~~~
\varphi_{\pi,\sigma}(u)=-(\rho/m_b)\varphi_{\pi,\sigma}'(1)\,,~~~
\varphi_{p}(u)=1,
\nonumber
\\
\varphi_\perp(1-u,0,u)= \widetilde{\varphi}_\perp(1-u,0,u) \simeq 
-(\rho/m_b)\widetilde{\varphi}'_{\perp}(0,0,1)\,,
\label{hql2}
\eea
where $ \widetilde{\varphi}'_{\perp}(0,0,1)=
[\partial\widetilde{\varphi}_{\perp}(1-u,0,u)/\partial u]_{u=1}$.
With these substitutions it is easy to reproduce, e.g. the leading $1/m_b^{3/2}$ 
behavior \cite{CZ}  of the $B\to \pi$ form factor from the sum rule (\ref{Bpilcsr}) 
at $m_b \to \infty$:
\be
f^+_{B\pi}(0) \Big |_{m_b\to \infty} =\frac{\hat{f}^+_{B\pi}(0)}{m_b^{3/2}}+O(1/m_b^{5/2})\,,
\ee
where \cite{KRW} 
\bea
\hat{f}^+_{B\pi}(0)=\frac{f_\pi}{2\hat{f}_B}e^{\overline{\Lambda}/\tau}
\int\limits_0^{2\omega_0} d\rho\; e^{-\frac{\rho}{2\tau}}
\left[-\rho\varphi_\pi'(1)+\mu_\pi\left(\varphi_p(1)-\frac{1}{6}\varphi_\sigma'(1)
\right)\right] 
\label{hatf}
\eea
is an effective, $m_b$-independent form factor.
Consequently, the heavy-quark mass limit for the factorizable amplitude 
obtained from Eq.~(\ref{fact}) is 
\be
A^{(O_1)}_E( \bar{B}^0_d \to \pi^+ \pi^-)\Big |_{m_b\to \infty}
= i\sqrt{m_b}f_\pi \hat{f}^+_{B\pi}(0).
\label{afact}
\ee
Applying Eqs.~(\ref{hql1}) and (\ref{hql2}) to the sum rule
(\ref{lcsr}) one obtains the limiting behavior for the separate
contributions:
\bea 
A^{(O_{8g})}_{hard}\Big |_{m_b\to \infty}\!\!\!&=&\!\!\!
i\frac{\alpha_sC_F}{2\pi}\sqrt{m_b}\Bigg(\frac{1}{4\pi^2f_\pi}\int\limits_0^{s_0^\pi}
ds~e^{-s/M_1^2}\Bigg)
\nonumber
\\
\!\!\!&\times&\!\!\! \Bigg(\frac{f_\pi}{2\hat{f}_B}e^{\overline{\Lambda}/\tau}
\int\limits_0^{2\omega_0} d\rho e^{-\frac{\rho}{2\tau}}
\left[-\rho\varphi_\pi'(1)+\mu_\pi\left(\frac32\varphi_p(1)-\frac{1}{12}
\varphi_\sigma'(1)\right)\right] \Bigg) \,,
\label{mlim1}
\\
A^{(O_{8g})}_{soft}\Big |_{m_b\to \infty}\!\!\!&=&\!\!\!i\sqrt{m_b}
\Bigg(\frac{1}{4\pi^2f_\pi}\int\limits_0^{s_0^\pi}
ds~e^{-s/M_1^2}\Bigg)
\Bigg(\frac{4f_\pi}{\hat{f}_B m_b^2}e^{\overline{\Lambda}/\tau}
\int\limits_0^{2\omega_0} d\rho e^{-\frac{\rho}{2\tau}}\rho
\,\widetilde{\varphi}'_{\perp}(0,0,1)
\Bigg)\,,
\label{mlim2}
\\
A^{(O_{8g})}_{\langle\bar{q}q\rangle}\Big |_{m_b\to \infty}\!\!\!&=&\!\!\!
i\frac{\alpha_sC_F}{3\pi}\sqrt{m_b}
\Bigg(\frac{-\langle \bar{q}q\rangle}{f_\pi m_b}\Bigg)
\nonumber
\\
\!\!\!&\times&\!\!\! \Bigg(\frac{f_\pi}{2\hat{f}_B}e^{\overline{\Lambda}/\tau}
\int\limits_0^{2\omega_0} d\rho e^{-\frac{\rho}{2\tau}}
\left[-\rho\varphi_\pi'(1)+\mu_\pi\left(\frac32\varphi_p(1)-\frac1{12}
\varphi_\sigma'(1)\right)\right] \Bigg) \,,
\label{mlim3}
\eea
where only the leading in $1/m_b$ term in each of the three
contributions is retained. We see that 
the hard-gluon contribution (\ref{mlim1}) has only
$O(\alpha_s)$ suppression with respect to the factorizable amplitude
(\ref{afact}). Note that in Eq.~(\ref{mlim1}) only the diagram in
Fig.~1a  contributes, because the twist 3 contribution of the Fig.~1b
diagram in Eq.~(\ref{Ahard}) has
an additional $1/m_b$ suppression.
Importantly, the soft gluon contribution (\ref{mlim2}), where the pion
DA is proportional to the dimensionful parameter $\delta_\pi^2$,  is
suppressed by two powers 
of $1/m_b$ . Finally, the quark-condensate contribution 
(\ref{mlim3}) is both $O(\alpha_s)$ and $O(1/m_b)$ suppressed. 
Thus, in the heavy-quark
mass limit the sum rule (\ref{lcsr}) behaves 
similar to the LCSR for the $B\to\pi$ form factor:
the twist 2 and 3 DA survive in the asymptotic limit,
whereas the higher-twist contributions 
are suppressed by powers of $1/m_b$.

Furthermore, we use the relation between the twist 3 DA: 
$\varphi'_\sigma(u)=6(1-2u)\varphi_p(u)$, 
which follows \cite{BF} from QCD equations of motion in the approximation 
where the quark-antiquark-gluon DA are neglected. Hence,
at $u\to 1$, $\varphi_p(1)=-\varphi'_\sigma(1)/6$, 
and the expression in the second
line in Eq.~(\ref{mlim1}) (and the same in Eq.~(\ref{mlim3})) is simply equal 
to $\hat{f}^+_{B\pi}(0)$ given by Eq.~(\ref{hatf}). 
As a result, at $m_b\to \infty$, 
the gluonic penguin amplitude is approximately 
equal to the heavy-quark limit (\ref{afact}) of the factorizable amplitude 
multiplied by an $O(\alpha_s)$ factor:
\bea
 A^{(O_{8g})}( \bar{B}^0_d \to \pi^+\pi^-)\Bigg|_{m_b\to \infty}=
 A^{(O_{8g})}_{hard}\Bigg|_{m_b\to \infty}=
i\frac{\alpha_sC_F}{2\pi}\sqrt{m_b}\Bigg(\frac{1}{4\pi^2f_\pi}\int\limits_0^{s_0^\pi}
ds~e^{-s/M_1^2}\Bigg)\hat{f}^+_{B\pi}(0)
\nonumber
\\
\simeq
\frac{\alpha_s C_F}{2\pi} A^{(O_1)}_E( \bar{B}^0_d \to \pi^+\pi^-)\Bigg|_{m_b\to \infty}\,,
\label{limit}
\eea
where the approximate relation (\ref{bracket}) is used.
Note that the above factorization 
will be violated by $O(1/m_b)$ corrections to Eq.~(\ref{mlim1}). The 
latter corrections 
can be easily obtained expanding Eq.~(\ref{Ahard}) to the next-to-leading 
order in the inverse heavy quark mass.   
Interestingly however, the quark-condensate contribution (\ref{mlim3}),
being $1/m_b$ suppressed with respect to Eq.~(\ref{mlim1}),
nevertheless reveals the same factorization property:
\be
A^{(O_{8g})}_{\langle\bar{q}q\rangle}\Big |_{m_b\to \infty}\!\!=\!
i\frac{\alpha_s C_F}{6\pi}\left(\frac{\mu_\pi}{m_b}\right)
\sqrt{m_b}f_\pi\hat{f}^+_{B\pi}(0)
\!=\!\frac{\alpha_s C_F}{6\pi}\left(\frac{\mu_\pi}{m_b}\right) 
A^{(O_1)}_E( \bar{B}^0_d \to \pi^+\pi^-)\Bigg|_{m_b\to \infty}\,,
\label{qqlimit}
\ee
where the PCAC relation $\mu_\pi=-2\langle\bar{q}q\rangle/f_\pi^2$ is
used  to replace the quark-condensate density by $\mu_\pi$.
Naturally, the soft-gluon contribution (\ref{mlim2}) also violates the
factorization relation (\ref{limit}).
Although this contribution is of order $1/m_b^2$ with respect to 
$A^{(O_{8g})}_{hard}\Big |_{m_b\to \infty}$
, it has no $\alpha_s$ suppression and therefore should be considered
separately from the $O(\alpha_s/m_b^2)$
corrections stemming from the heavy-mass expansion 
of hard-gluon and quark-condensate terms in LCSR beyond $O(1/m_b)$. A direct evaluation 
of the latter corrections from Eqs.~(\ref{Ahard}) and (\ref{Aqq})
will, however, be incomplete because, in deriving LCSR, we have
already neglected small (calculable) $s_0^\pi/m_B^2$ terms  of the same order.

\section{Numerical results}
\begin{figure}[t]
\begin{center}
\includegraphics[width=0.8\textwidth]{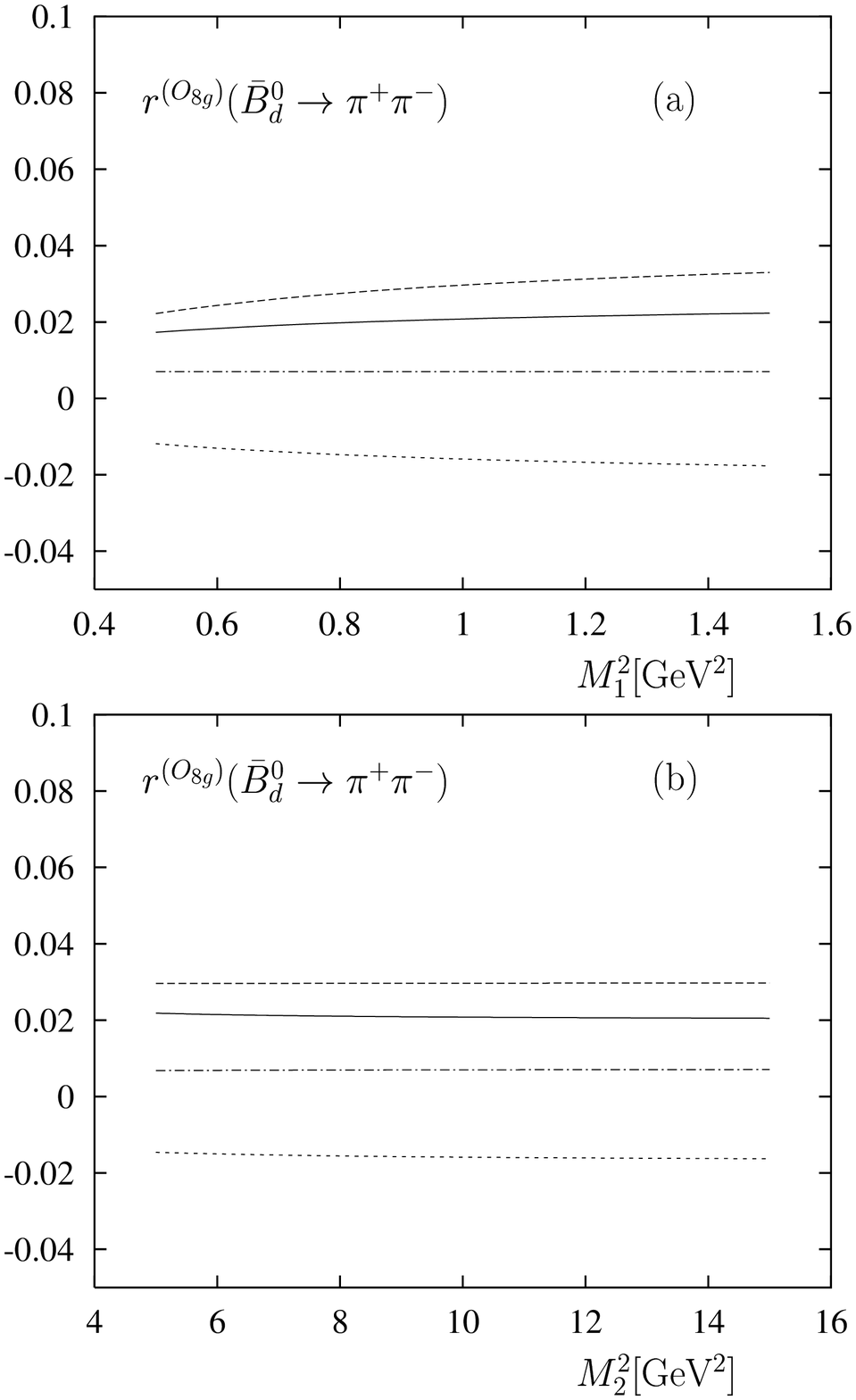}
\end{center}
\caption{
{\it The ratio of the $\langle \pi^+\pi^-|O_{8g}|\bar{B}^0_d\rangle $ 
and $\langle \pi^+\pi^-|O_{1}|\bar{B}^0_d\rangle $ matrix elements
calculated from LCSR (solid line) 
as a function of the Borel parameters $M_1$  in the pion channel
(a)  and $M_2$ in the $B$ channel (b). 
Long-dashed, short-dashed and dash-dotted lines 
are the hard-gluon, soft-gluon  and quark-condensate contributions
to the sum rule, respectively.}
\label{fig:fig3}}
\end{figure}
Let us return to the sum rule (\ref{lcsr}) at the finite $b$-quark mass
and use it to obtain  a numerical estimate for the ratio
\be
r^{(O_{8g})}( \bar{B}^0_d \to \pi^+ \pi^-)= A^{(O_{8g})}( \bar{B}^0_d \to \pi^+ \pi^-)/
A^{(O_1)}_E( \bar{B}^0_d \to \pi^+ \pi^-)\,,
\label{ratio}
\ee
which determines (up to the known Wilson coefficient $c_{8g}$) 
the gluonic-penguin correction to the factorizable $B\to \pi\pi$ decay amplitude.  
The ratio (\ref{ratio})
has an important advantage of being less sensitive to 
the input parameters than the individual matrix elements $A^{(O_8)}$ and
$A^{(O_1)}_E$. Moreover, $r^{(O_{8g})}$ is independent 
of $f_B$, since the latter cancels in the ratio of the sum rules  (\ref{lcsr}) and (\ref{Bpilcsr}).

The parameters for the pion channel are $f_\pi=132$ MeV and $s_0^\pi=0.7$ GeV$^2$,
the latter determined from the two-point SVZ sum rule \cite{SVZ} for $f_\pi$. 
The corresponding Borel parameter interval is $M^2= 0.5\,\mbox{-}1.5 $ GeV$^2$,
accommodating also the range used in the LCSR for the
pion form factor \cite{BKM}. 

The inputs for the $B$ channel are taken from the earlier analysis 
of LCSR (\ref{Bpilcsr}) for the $B\to \pi$ form factor in Refs.~\cite{Bpi,KRWY}. 
In particular, to be consistent with the latter sum rule, we identify
$m_b$ with the one-loop pole mass and adopt the interval $m_b=4.7 \pm
0.1$ GeV which is in accordance with the 
most recent average of the $\overline{\mbox{MS}}$ mass $\bar{m}_b(\bar{m}_b)=4.24\pm
0.11$ GeV \cite{ElKLuke} obtained from  various model-independent
determinations (including QCD sum rules for the $\Upsilon$ system). 
Fixing the $m_b$ interval one determines the corresponding duality
threshold  $s_0^B= 35\mp 2$ GeV$^2$ 
from the QCD sum rule for $f_B$ taken with $O(\alpha_s)$ accuracy.
Furthermore we take the range $M_2^2=8\,\mbox{-} 12$ GeV$^2$ for the second Borel
parameter and $\mu_b= \sqrt{m_B^2-m_b^2}\simeq 2.4$ GeV for the related scale 
at which the pion DA are normalized. The same scale is adopted for the 
$\alpha_s$ normalization. For the latter we also use the two-loop
running with $\bar\Lambda^{(4)}=280$ MeV. However, one should have in
mind that in the new sum rule 
(\ref{lcsr}) two independent Borel parameters $M_1$ and $M_2$ are
present, and the scale $M_1\sim $ 1 GeV is independent of $m_b$.
To investigate the sensitivity to the normalization scale, we will vary
it in a rather wide interval, between   $\mu_b/2$ and $2\mu_b$.
For the  quark condensate density we take  
$\langle \bar{q}q \rangle(1 \mbox{GeV})=(-240\pm 10 ~\mbox{MeV})^3$,
or, equivalently, $\mu_\pi(1 \mbox{GeV})= 1.59 \pm 0.2 $ GeV. The 
normalization parameter of the twist 4 DA 
$\delta_\pi^2(1 \mbox{GeV})=0.17\pm 0.05 $ GeV$^2$ is determined from
the two-point QCD sum rules \cite{delta} (see also Ref.~\cite{BK}). 

For our initial numerical illustration all pion light-cone DA are taken
in the asymptotic form, 
that is, the coefficients $a_2,f_{3\pi},\omega_{3\pi},\epsilon_\pi$
are put to zero in the expressions for DA presented in 
the appendix. The numerical results for the ratio $r^{(O_{8g})}$, together with the 
hard-, soft-gluon and quark-condensate contributions, calculated  
at the central values of the input parameters and for the
asymptotic form of the pion DA are plotted in Fig.~3 as a function of
Borel variables. We observe a nice stability in both $M_1,M_2$. 
The most important feature of our calculation which is 
clearly seen in Fig.~3 is that the soft-gluon part
of the gluonic penguin amplitude, while being suppressed by two powers
of $m_b$, numerically amounts to about 50\%  of the hard-gluon contribution and 
has an opposite sign. A 
relatively large magnitude of the soft-gluon term (\ref{Asoft}) can be traced 
back to the large numerical factor of 20 originating from the contributing twist-4
DA and compensating the suppression factor $\delta_\pi^2/m_b^2$. 
The quark-condensate contribution, on the other hand, is subleading
also numerically, not larger than  30\% of the hard-gluon contribution. 

To take into account deviations from the
asymptotic form, we have  repeated our calculation   
with the nonasymptotic part of the pion DA
taken to the next-to-leading order in the conformal spin \cite{BF}.
In this case we adopt in the pion twist-2 DA $a_2(1 \mbox{GeV})=0.24\pm 0.14\pm 0.08 $
obtained \cite{BK} by fitting the LCSR for the
pion form factor to the experimental data. The relevant parameters in
the twist 3,4 DA have been estimated 
at the scale of $\sim 1$ GeV  using two-point QCD sum rules: 
$f_{3\pi}=0.0035$, $\omega_{3\pi}=-2.88$  \cite{ZZC} and $\epsilon_\pi=0.5$ \cite{BF}  . 
These estimates have an uncertainty of $\pm 30\%$. The corresponding LO anomalous
dimensions can be found  in the appendix.  

The difference between the values of $r^{(O_{8g})}$ obtained with  
nonasymptotic and asymptotic DA is inessential for the
hard-gluon and quark-condensate contributions. On the contrary, in 
the soft-gluon term,  the effect of the nonasymptotic part determined by the parameter
$\epsilon_\pi$ is substantial. The reason is that in the soft-gluon
term described by the diagram in Fig.~2a the gluon carries the dominant
fraction $\alpha_3=u$ ($u_{0}^B<u<1$)  of the pion momentum,
the $d$-quark carries $\alpha_1=1-u$, whereas  the $\bar{u}$-quark momentum fraction
is restricted to small values: $\alpha_2\sim
s_0^\pi/m_B^2$. The integration 
over the variables
$\alpha_i$ in this subspace of the integration region $\sum_i\alpha_i=1$ 
is strongly influenced by the presence of the second polynomial in the
conformal expansion of the twist 4 DA $\varphi_\perp(\alpha_i)$ and
$\tilde{\varphi}_\perp(\alpha_i)$. As a result, the soft-gluon
contribution, being $-50\%$ of the hard-gluon part 
in the case of asymptotic DA, drops to $(-10\%) \,\mbox{-} (+20\%)$  at
$\epsilon_\pi=0.35\,\mbox{-} 0.65$.

Finally, we get the following 
numerical estimate for the ratio of the gluonic-penguin and factorizable
amplitudes including nonasymptotic effects:
\be
r^{(O_{8g})}(\bar{B}^0_d\to \pi^+\pi^-)= 0.035 \pm 0.015.
\label{number}
\ee
In obtaining the above  range, the uncertainties caused by the variation of all parameters
within their allowed intervals are added linearly. The uncertainty 
in Eq.~(\ref{number}) is mainly due to the sensitivity to the 
nonperturbative parameters
$\epsilon_\pi$, $\langle \bar{q}q \rangle $ and $\delta^2_\pi$, whereas the choice of
the $b$-quark mass, Borel parameters and $a_2$ has a smaller impact.   
There is an additional uncertainty of approximately $\pm 20\%$ from
varying the normalization scale in the limits $\mu_b/2 ~\mbox{-} \,2 \mu_b$. 
The LCSR prediction for the $B\to \pi$ form factor with the
same input and the same conservative procedure of estimating the uncertainty is 
$f^+_{B\pi}=0.28 \pm 0.06$.

One can also investigate the quality of the $m_b \to \infty$ limit for
the ratio $r^{(O_8g)}$. Dividing Eq.~(\ref{afact}) by Eq.~(\ref{limit}) we
find that, within uncertainties of our input, $r^{(O_{8g})}$ varies between 0.6
and 1.1 of its heavy quark limit. 
Future improvements of nonperturbative 
parameters determining the vacuum-pion matrix elements of twist 3,4 will
allow to decrease the uncertainty of our numerical estimates, in
particular, of the subleading in $1/m_b$ terms.  

\section{Gluonic penguins and the decay amplitude} 

We are now in a position to assess  the role of 
the gluonic-penguin matrix element in $\bar{B^0} \to \pi^+\pi^-$. 
The decay amplitude can be represented in the following form:
\bea
{\cal A}(\bar{B}^0_d\to \pi^+\pi^-) &\equiv& \langle \pi^+\pi^-|H_W| \bar{B^0_d}\rangle
\nonumber
\\
&=&i\frac{G_F}{\sqrt{2}} f_\pi f_{B\pi}^+(0)~m_B^2 \Bigg\{\lambda_u \Bigg[c_1(\mu) +\frac{c_2(\mu)}{3}
+
2c_2(\mu)r_{E}^{(\widetilde{O}_1)}
(\bar{B}^0_d \to \pi^+ \pi^-)\Bigg]
\nonumber
\\
&+&... +\lambda_t c_{8g}(\mu) r^{(O_{8g})}( \bar{B}^0_d \to \pi^+ \pi^-)
\Bigg\}\,,
\label{ampl}
\eea
where we denote by ellipses all terms that are not relevant 
for the present discussion: the hadronic matrix elements 
of $O_{1,2}$ in the annihilation, penguin and penguin-annihilation 
topologies; the contributions generated by current-current operators
with $c$-quarks (``charming penguins'') and, finally, all effects of  
quark- and electroweak-penguin operators (for a model-independent
classification of all contributions and topologies see
Ref.~\cite{BurasSilv}). 
The second line in Eq.~(\ref{ampl}) contains the factorizable amplitude 
and the nonfactorizable correction due to the operator
$\widetilde{O}_1$. The latter is parametrized by the ratio 
\be
r_{E}^{(\widetilde{O}_{1})}( \bar{B}^0_d \to \pi^+ \pi^-)
\equiv  A^{(\widetilde{O}_{1})}_{E}( \bar{B}^0_d \to \pi^+ \pi^-)/
A^{(O_1)}_E( \bar{B}^0_d \to \pi^+ \pi^-)\,,
\ee
where   $A^{(\widetilde{O}_{1})}_{E}$ is the $\bar{B}^0_d \to \pi^+
\pi^-$ matrix element of $\widetilde{O}_1$ in the emission
topology. Finally, in the third line in Eq.~(\ref{ampl})
we added the gluonic penguin correction calculated above.

The matrix element of $\widetilde{O}_1$ contains the hard $O(\alpha_s)$ 
and soft $O(1/m_b)$ contributions, so that 
\be
r_E^{(\widetilde{O}_{1})}( \bar{B}^0_d \to \pi^+ \pi^-) =
r_E^{(\widetilde{O}_{1},~hard)}( \bar{B}^0_d \to \pi^+ \pi^-) 
+\frac{\lambda_E(\bar{B}^0_d \to \pi^+ \pi^-)}{m_B} \,, 
\ee
where the soft contribution was calculated
from LCSR in  Ref.~\cite{AK} with an estimate 
$\lambda_E(\bar{B}^0_d \to \pi^+ \pi^-)= 0.1\pm 0.05 $ GeV. 
The hard nonfactorizable correction 
$r_E^{(\widetilde{O}_1,hard)}$ is also accessible with LCSR 
but demands two-loop calculation. Following Ref.~\cite{AK} we simply 
use the QCD factorization estimate: 
$r_E^{(\widetilde{O}_{1},hard)}( \bar{B}^0_d \to \pi^+ \pi^-)=
\alpha_s C_FF/(8\pi N_c)$, where the expression for $F$ is given in Ref.~\cite{BBNS}.
Since we are only interested in an order-of-magnitude estimate, we
neglect the twist-3 part of $r_E^{(\widetilde{O}_{1},hard)}$ which in  
QCD factorization \cite{BBNS} has to be regularized introducing some additional parameter.
For consistency we use the same normalization scale $\mu_b$ and twist-2 pion DA as in LCSR. We obtain:  
\be
r_E^{(\widetilde{O}_{1})}( \bar{B}^0_d \to \pi^+ \pi^-)= -[(0.002 \,\mbox{-} 0.016)+0.045i]
+[0.01\,\mbox{-} 0.03]\,,
\label{nonfact1}
\ee
where the first and second brackets contain the
hard-gluon (QCD factorization) and soft-gluon (LCSR) contributions.
The relatively large uncertainty of the hard-gluon part  is due to the
variation of $a_2(1 \,\mbox{GeV})$ in the adopted range 0.02-0.46. 
Substituting the estimates (\ref{number}) and (\ref{nonfact1}) 
in Eq.~(\ref{ampl})  
and using also  the LCSR estimate for the $B\to \pi$ form factor we obtain:
\bea
&&{\cal A}(\bar{B}^0_d\to \pi^+\pi^-)
= i\frac{G_F}{\sqrt{2}} f_\pi~m_B^2 (0.28 \pm 0.05)
\Bigg\{\lambda_u \Big(1.03 
\nonumber
\\
&& +[0.001\,\mbox{-} 0.009 +0.025i] -[0.005\,\mbox{-} 0.015]\Big)
+...-\lambda_t[0.003\,\mbox{-} 0.008]\Bigg\}\,,
\label{amplnum}
\eea
where, for consistency, the Wilson coefficients \cite{Hrev} are taken 
at the scale $\mu_b$. We use  $c_1(\mu_b)=1.124$ and $c_2(\mu_b)=-0.272$
in NLO and in the NDR scheme and, correspondingly,
$c_{8g}(\mu_b)=-0.166$ in LO.  
In Eq.~(\ref{amplnum}) the first (second) bracket in the term
proportional to $\lambda_u$ is due to the hard (soft) parts of $r_E^{(\widetilde{O}_{1})}$.
Note that some of the uncertainties in the $B\to \pi$ form factor and in
the nonfactorizable contributions are correlated. 

Both nonfactorizable corrections in Eq.~(\ref{amplnum})
turn out to be very small, not exceeding  a one-percent
level. However, before one can claim that $\bar{B}^0_d \to \pi^+ \pi^-$ 
has a predominantly factorizable amplitude, it is necessary to calculate 
all remaining contributions, having in mind that the sum of all nonfactorizable
contributions may accumulate into a noticeable effect. As both analyses
of Ref.~\cite{AK} and this paper
show, in this calculation the soft-gluon effects
are indispensable. 

Note that the situation with the colour-suppressed 
$\bar{B}^0_d \to \pi^0 \pi^0$ channel 
is quite different. Here the factorizable part is proportional to
the small coefficient
$c_2(\mu_b)+c_1(\mu_b)/3\simeq 0.1$, therefore the contribution
of the $\widetilde{O}_1$ matrix element multiplied by $2c_1$ is equally 
important. Also the gluonic-penguin correction in
this channel can reach a few \% of the total amplitude.
One may expect that in some other penguin-dominated channels 
of charmless $B$ decays, such as $B\to K\pi$,  the effects of gluonic
penguins may even be more important.

\section{Comparison with QCD factorization}

It is interesting to compare our estimate of the hadronic 
matrix element  $\langle \pi\pi |O_{8g}| B \rangle $ 
expressed via the ratio $r^{(O_{8g})}$ with the same ratio 
derived from QCD factorization. We can 
only compare the $O(\alpha_s)$ contributions 
in our calculation because, in the QCD factorization scheme, the
soft-gluon effects have not yet been investigated quantitatively. 

To simplify the comparison, we set to zero all Wilson coefficients in $H_W$ 
except $c_{8g}$.
In this case, the hard-scattering kernel in $O(\alpha_s)$ calculated 
within QCD factorization approach contains a single diagram 
with a hard gluon emitted by the $O_{8g}$ vertex and converted
into a quark-antiquark pair. This diagram (the sixth
one in Fig.~2 of the second paper in Ref.~\cite{BBNS}) is simply reproduced  
from the one in Fig.~1a of this paper if one puts on shell the quarks
emitted by the $B$-meson and pion currents. 
With the same procedure of putting on shell the quark
lines applied to the Fig.~1b diagram, 
one gets a diagram which  describes a hard-gluon exchange between 
the gluonic-penguin vertex and the spectator quark in the $B$-meson, with a subsequent 
hadronization of the light-quark pair into a pion pair. The
long-distance part of this process can be described by a 
two-pion DA, an object different from the one-pion DA used before.
In the framework of QCD factorization this particular diagram 
has not been taken into account. A convincing reason for that 
is that the corresponding contribution is $1/m_b$ suppressed, which is also confirmed by the
heavy-mass expansion of LCSR discussed above. 
We conclude, that, apart from soft-gluon effects, at finite $m_b$ there are certain
hard-gluon effects in LCSR, such as the contribution of the Fig.~1b diagram, 
which do not appear in the QCD factorization. 

After this qualitative discussion we 
turn to the quantitative comparison with the gluonic penguin
contribution in the QCD factorization. 
In our toy scenario with only $c_{8g}\neq 0$ 
it is easy to extract this contribution from the expression for the effective 
coefficients $a_i$ obtained in Refs.~\cite{BBNS} for $B\to \pi\pi$.
The only effective coefficients where   $c_{8g}$ contributes 
are $a_4$ and $a_6$, so that the ratio $r^{(O_{8g})}$ in the QCD factorization scheme is :
\be
r^{(O_{8g})}(\bar{B}^0_d\to \pi^+\pi^-)_{QCD fact.}= \frac{\alpha_sC_F}{2\pi N_c}\left( \int_0^1du
\frac{\varphi_\pi(u)}{1-u} + \frac{2\mu_\pi}{m_b}\right)\,,
\label{bbns}
\ee 
together with the $O(\mu_\pi/m_b)$ correction which is the only $1/m_b$
effect retained in QCD factorization \cite{BBNS}
because of the large numerical value of $\mu_\pi$.
At  $m_b\to \infty$  the above ratio, for the 
asymptotic DA $\varphi_\pi=6u(1-u)$, coincides with   
the heavy-quark mass limit (\ref{limit}) obtained from
LCSR, if in the latter the approximation (\ref{bracket})
is used. 
Thus, the LCSR prediction for the gluonic-penguin matrix 
element reproduces the QCD factorization result for the same matrix
element in the limit of infinitely heavy $b$ quark. 

The two methods, however, substantially differ 
in subleading terms. This difference manifests itself already 
at the $O(\alpha_s/m_b)$ level. In addition to the penguin-spectator exchange
effect corresponding to the diagram in Fig.~2b and discussed above,
there is a factor of 2 difference between the second, subleading term in Eq.~(\ref{bbns}) 
and the heavy quark mass limit (\ref{qqlimit}) 
of the quark-condensate contribution  in LCSR. 
Furthermore, the fact that the soft gluon corrections 
in LCSR are suppressed by two powers of $1/m_b$ is in a general 
agreement with the expectations of QCD factorization.
However, as we already noted, this effect which is absent in 
the QCD factorization prediction, turns out to be 
essential at finite $m_b$.

To complete our comparison with other methods let us
parenthetically note that LCSR predictions 
do not support the PQCD  approach \cite{pqcd} to charmless
$B$ decays. In particular, according to  LCSR and contrary to PQCD, 
the soft contributions dominate in the $B\to \pi$ form factor.

\section{Conclusion}
In this paper, we have presented the  first calculation of 
the hadronic $ B \to \pi \pi$ matrix element of the gluonic penguin operator
$O_{8g}$ using QCD LCSR.  This matrix element is a very suitable study object 
for the sum rule approach because both hard- and soft-gluon contributions 
are calculable within one procedure using the same input. 
The results obtained for $\langle\pi^+ \pi^-|O_{8g}|\bar{B}_d^0\rangle$
from LCSR  clearly indicate that
at finite $m_b$ the soft-gluon and other $1/m_b$ suppressed effects are
important motivating
further investigation in this direction, in particular, calculating the
penguin matrix elements of current-current operators. 
Importantly, in the heavy-quark mass limit the LCSR prediction is consistent 
with the QCD factorization. Similar to the 
nonfactorizable emission, the gluonic-penguin contributions to $B\to \pi\pi$ are suppressed
either by $\alpha_s$ or by powers of $1/m_b$.
On the phenomenological side, our result clearly indicates 
that  gluonic penguins play an insignificant role in the  $ \bar{B}^0_d \to \pi^+ \pi^-$
decay amplitude but become noticeable in  $ \bar{B}^0_d \to \pi^0 \pi^0$. Our prediction can also be used 
in the future analyses of penguin-dominated $B\to K\pi$ modes.

\subsection*{Acknowledgments}

This work is supported by the DFG Forschergruppe ``Quantenfeldtheorie,
Computeralgebra und Monte Carlo Simulationen'', by the German Ministry
for Education and Research (BMBF),  and 
by the KBN grant 5P03B09320. P.U. acknowledges the support of the Humboldt
Foundation. 

\app
\section*{Appendix: Vacuum-pion matrix elements and 
distribution amplitudes }
The following convenient decomposition of the 
vacuum-pion matrix element near the light-cone is used in 
the calculation of the correlation function (\ref{corr}):
\bea
\langle 0 |\bar{u}^{i}_\xi(x_2) d^j_{\omega}(x_1)|\pi^-(q)\rangle
=-\frac{i\delta^{ij}}{12}f_\pi\int\limits_0^1 \!du\, e^{-iuqx_1-i\bar{u}qx_2}
\Big[(\not\!q\gamma_5)_{\omega\xi}\varphi_\pi(u)
\nonumber
\\
+(\gamma_5)_{\omega\xi}\mu_\pi\varphi_p(u)-
\frac{1}{6}(\sigma_{\beta\tau}\gamma_5)_{\omega\xi}q_\beta(x_1-x_2)_\tau
\mu_\pi\varphi_\sigma(u)\Big]+ \mbox{twist $\geq$ 4}\,,
\eea
where the quark fields are taken near the light-cone ($x_i=u_ix$, $x^2=0$),
 $i,j=1,2,3$ and $\xi,\omega=1\,\mbox{-}4$ are the quark color and spinor indices,
respectively. Only the twist-2 DA $\varphi_\pi$ and the twist 3 DA 
$\varphi_p$ and $\varphi_\sigma$ are retained. The familiar definitions of 
these DA are obtained by multiplying both parts 
of this equation by the corresponding combinations 
of $\gamma$ matrices and taking Dirac and color traces.
For the twist-2 pion DA, in the next-to-leading 
accuracy in the conformal spin \cite{BF} adopted in this paper, 
only the first nonasymptotic term is retained
in the expansion in Gegenbauer polynomials $C_{2n}^{3/2}$: 
\begin{equation}
\varphi_\pi(u,\mu ) = 6 u \bar u \left[ 1 + 
a_{2}(\mu) C_{2}^{3/2}(u - \bar u) \right]\,.
\label{phipi}
\end{equation}
The LO scale-dependence of the nonasymptotic part 
is determined by 
\begin{equation} 
a_{2}(\mu_2)=\left[L(\mu_2,\mu_1)\right]^{\gamma_2^{(0)}/\beta_0}
a_{2}(\mu_1)~,
\label{anom}
\end{equation}
where $L(\mu_2,\mu_1)=\alpha_s(\mu_2)/\alpha_s(\mu_1)$, 
$\beta_0=11- \frac23 N_F$, and $\gamma_{2}^{(0)}=50/9$.

The analogous decomposition for the quark-antiquark-gluon matrix element
in terms of the pion three-particle DA reads :
\bea
\langle 0 |\bar{u}^{i}_\xi(x_2) g_sG^a_{\mu\nu}(x_3) 
d^j_{\omega}(x_1)|\pi^-(q)\rangle
=\frac{\lambda^a_{ji}}{32}\int {\cal D}\alpha_ie^{-iq(\alpha_1 x_1+\alpha_2 x_2+\alpha_3x_3)}
\nonumber
\\
\times\Bigg[if_{3\pi}(\sigma_{\lambda\rho}\gamma_5)_{\omega\xi}
(q_\mu q_\lambda g_{\nu\rho}-
q_\nu q_\lambda g_{\mu\rho})\varphi_{3\pi}(\alpha_i)
\nonumber
\\
-f_\pi(\gamma_\lambda\gamma_5)_{\omega\xi}\Big\{(q_\nu 
g_{\mu\lambda}-q_\mu g_{\nu\lambda})\varphi_\perp(\alpha_i)
+\frac{q_\lambda(q_\mu x_\nu-q_\nu x_\mu)}{(q\cdot x)}
\left(\varphi_\parallel(\alpha_i)+\varphi_\perp(\alpha_i)\right)
\Big\}
\nonumber
\\
+\frac{if_\pi}2\epsilon_{\mu\nu\delta\rho}(\gamma_\lambda)_{\omega\xi}
\Big\{(q^\rho g^{\delta\lambda}-q^\delta g^{\rho\lambda})
\widetilde{\varphi}_\perp(\alpha_i)+  
\frac{q_\lambda(q^\delta x^\rho-q^\rho x^\delta)}{(q\cdot x)}
\left(\widetilde{\varphi}_\parallel(\alpha_i)+
\widetilde{\varphi}_\perp(\alpha_i)\right)\Big\}\Bigg]\,.
\label{decomp2}
\eea
In the above
\be
\varphi_{3\pi}=360\alpha_1\alpha_2\alpha_3^2\left[1+
\frac{\omega_{3\pi}}{2}(7\alpha_3-3)\right]
\ee
is the twist-3 quark-antiquark-gluon DA,
with nonperturbative parameters $f_{3\pi}$ and 
$\epsilon_\pi$ defined via matrix
elements of the following local operators:
\bea 
\langle 0 |\bar{u}
\sigma_{\mu\nu}\gamma_5 G_{\alpha\beta}d
| \pi^-(q) \rangle 
=if_{3\pi}\Big[(q_\alpha q_\mu g_{\beta\nu}-q_\beta q_\mu g_{\alpha\nu})
-(q_\alpha q_\nu g_{\beta\mu}-q_\beta q_\nu g_{\alpha\mu})\Big]\,,
\label{tw3matr}
\eea
\be
\langle 0 |\bar{u}\sigma_{\mu\lambda}\gamma_5
[D_\beta,G_{\alpha\lambda}] d-
\frac37 \partial_\beta\bar{u}\sigma_{\mu\lambda}\gamma_5
G_{\alpha\lambda}d
|\pi^-(q)\rangle= \frac{3}{14}f_{3\pi}\omega_{3\pi}q_\alpha q_\beta q_\mu\,. 
\ee
The scale dependence of the twist 3 parameters is given by:
\begin{eqnarray}
\mu_{\pi}(\mu_2) = \left[L(\mu_2,\mu_1)\right]
^{-\frac{4}{\beta_0}}\mu_{\pi}(\mu_1)\,,~~~
f_{3\pi}(\mu_2) = \left[L(\mu_2,\mu_1)\right]^
{\frac{1}{\beta_0}\left(\frac{7 C_F}{3}+3\right)}
f_{3\pi}(\mu_1)\,,
\\
(f_{3\pi}\omega_{3\pi})(\mu_2) = 
\left[L(\mu_2,\mu_1)\right]^{\frac{1}{\beta_0}\left(\frac{7 C_F}{6}+10\right)}
(f_{3\pi}\omega_{3\pi})(\mu_1)\,,
\end{eqnarray}
The corresponding expressions for the twist-3 quark-antiquark DA are:
\bea
\varphi_p(u)=1+30\frac{f_{3\pi}}{\mu_\pi f_\pi} C_2^{1/2}(u-\bar u)-
3\frac{f_{3\pi}\omega_{3\pi}}{\mu_\pi f_\pi} C_4^{1/2}(u-\bar{u}),
\nonumber
\\
\varphi_\sigma(u)=6u(1-u)\left(1+5\frac{f_{3\pi}}{\mu_\pi f_\pi}
\left(1-\frac{\omega_{3\pi}}{10}\right)C_2^{3/2}(u-\bar u)\right).
\eea

Finally, the expressions for the four twist-4 DA 
entering the decomposition (\ref{decomp2})
are:
\bea
&&\varphi_{\parallel}(\alpha_i)=
120\delta^2_\pi\epsilon_\pi(\alpha_1-\alpha_2)\alpha_1\alpha_2\alpha_3,
\nonumber \\
&&
\varphi_{\perp }(\alpha_i)=30\delta^2_\pi(\alpha_1-\alpha_2)\alpha_3^2
\left[\frac{1}3 +2\epsilon_\pi(1-2\alpha_3)\right]\,,
\nonumber \\
&&\widetilde \varphi_{\parallel } (\alpha_i) 
=- 120 \delta_\pi^2 \alpha_1\alpha_2 \alpha_3\left[\frac{1}{3}+\epsilon_\pi(1-3\alpha_3)\right],
\nonumber \\
&&
\widetilde \varphi_\perp(\alpha_i)=30\delta_\pi^2 
\alpha_3^2(1-\alpha_3)\left[\frac{1}3 +2\epsilon_\pi(1-2\alpha_3)\right]\,,
\label{tw43p}
\end{eqnarray}
where the nonperturbative parameters $\delta^2_\pi$ and $\epsilon_\pi$ are 
defined as
\begin{equation}
\langle 0 |\bar{u}\widetilde{G}_{\alpha\mu}\gamma^\alpha d
|\pi^-(q) \rangle=-i\delta^2_\pi f_\pi q_\mu \,,
\label{delta1}
\end{equation}
and (up to twist 5 corrections):
\be
\langle 0 |\bar{u}[D_\mu,\widetilde{G}_{\nu\xi}]\gamma^\xi d-
\frac49\partial_\mu\bar{u}\widetilde{G}_{\nu\xi}\gamma^\xi d
|\pi^-(q)\rangle= -\frac{8}{21}f_\pi\delta_\pi^2\epsilon_\pi q_\mu 
q_\nu \, ,
\ee
with the scale-dependence:
\be
\delta^2_\pi(\mu_2) = \left[L(\mu_2,\mu_1)\right]
^{\frac{8C_F}{3\beta_0}}\delta^2_\pi(\mu_1)\,,~~
(\delta_\pi^2\epsilon_\pi)(\mu_2) = 
\left[L(\mu_2,\mu_1)\right]
^{\frac{10}{\beta_0}}(\delta^2_\pi\epsilon_\pi)(\mu_1)\,.
\ee

\appende


\begin{thebibliography}{99}

\bibitem{Bpipiexp} 
D.~Cronin-Hennessy {\it et al.}  [CLEO Collaboration],
Phys.\ Rev.\ Lett.\  {\bf 85} (2000) 515;
B.~Aubert  [BABAR Collaboration],
hep-ex/0205082;\\
B.~C.~Casey {\it et al.}  [Belle Collaboration],
arXiv:hep-ex/0207090.


\bibitem{Fleischer}
R.~Fleischer,
hep-ph/0207108.

\bibitem{BBNS}
M.~Beneke, G.~Buchalla, M.~Neubert and C.~T.~Sachrajda,
Phys.\ Rev.\ Lett.\  {\bf 83} (1999) 1914;
Nucl.\ Phys.\ B {\bf 606} (2001) 245.



\bibitem{charmpeng}
M.~Ciuchini, E.~Franco, G.~Martinelli, M.~Pierini and L.~Silvestrini,
Phys.\ Lett.\ B {\bf 515} (2001) 33.

\bibitem{lcsr}
I.~I.~Balitsky, V.~M.~Braun and A.~V.~Kolesnichenko,
 Nucl.\ Phys.\  {\bf B312} (1989) 509.

\bibitem{BF1}
V.~M.~Braun and I.~E.~Filyanov,
 Z.\ Phys.\  {\bf C44} (1989) 157.

\bibitem{CZ}
V.~L.~Chernyak and I.~R.~Zhitnitsky,
Nucl.\ Phys.\  {\bf B345} (1990) 137.


\bibitem{AK}
A.~Khodjamirian,
Nucl.\ Phys.\  {\bf B605} (2001) 558.

\bibitem{Bpi}
V.~M.~Belyaev, A.~Khodjamirian and R.~R\"uckl,
Z.\ Phys.\  {\bf C60} (1993) 349,
V.~M.~Belyaev, V.~M.~Braun, A.~Khodjamirian and R.~Ruckl,
Phys.\ Rev.\ D {\bf 51} (1995) 6177.


\bibitem{KRWY}
A.~Khodjamirian, R.~R\"uckl, S.~Weinzierl and O.~Yakovlev,
Phys.\ Lett.\  {\bf B410} (1997) 275.
\bibitem{BBB}
E.~Bagan, P.~Ball and V.~M.~Braun,
 Phys.\ Lett.\  {\bf B417} (1998) 154.

\bibitem{BallZ}
P.~Ball and R.~Zwicky,
JHEP {\bf 0110} (2001) 019.


\bibitem{FORM} 
J.~A.~Vermaseren,
math-ph/0010025.


\bibitem{BKM}
V.~M.~Braun, A.~Khodjamirian and M.~Maul,
Phys.\ Rev.\ D {\bf 61} (2000) 073004.



\bibitem{SVZ}
M.~A.~Shifman, A.~I.~Vainshtein and V.~I.~Zakharov,
Nucl.\ Phys.\ B {\bf 147} (1979) 385,
448.



\bibitem{KR}
A.~Khodjamirian and R.~Ruckl,
in {\em Heavy Flavors}, 2nd edition, eds., A.J. Buras and M. Lindner,
World Scientific (1998), p. 345, hep-ph/9801443.


\bibitem{KRW}
A.~Khodjamirian, R.~Ruckl and C.~W.~Winhart,
Phys.\ Rev.\ D {\bf 58} (1998) 054013.



\bibitem{ElKLuke}
A.~X.~El-Khadra and M.~Luke,
hep-ph/0208114.




\bibitem{delta}
V.L.~Chernyak, A.R.~Zhitnitsky and I.R.~Zhitnitsky;
Sov. J. Nucl. Phys. {\bf 38} (1983) 645;\\
V.A.~Novikov, M.A.~Shifman, A.I.~Vainshtein, M.B.~Voloshin and V.I.~Zakharov,
Nucl. Phys. {\bf B237} (1984) 525.



\bibitem{BK}
J.~Bijnens and A.~Khodjamirian,
hep-ph/0206252, to be published in Eur.~Phys.~J. C.


\bibitem{BF}
V.~M.~Braun and I.~E.~Filyanov,
Z.\ Phys.\  {\bf C48} (1990) 239;\\
P.~Ball,
JHEP {\bf 01} (1999) 010.


\bibitem{ZZC}
A.~R.~Zhitnitsky, I.~R.~Zhitnitsky and V.~L.~Chernyak,
Sov.\ J.\ Nucl.\ Phys.\  {\bf 41} (1985) 284.


\bibitem{BurasSilv}
A.~J.~Buras and L.~Silvestrini,
Nucl.\ Phys.\  {\bf B569} (2000) 3.




\bibitem{Hrev}
G.~Buchalla, A.~J.~Buras and M.~E.~Lautenbacher,
Rev.\ Mod.\ Phys.\  {\bf 68} (1996) 1125.






\bibitem{pqcd}  
Y.~Y.~Keum, H.~n.~Li and A.~I.~Sanda,
Phys.\ Rev.\ D {\bf 63} (2001) 054008,\\
for a recent review see 
H.~n.~Li,
hep-ph/0210198.


\end{thebibliography}
\end{document}